\documentstyle[preprint,pra,eqsecnum,aps]{revtex}
\tightenlines
\begin{document}
\draft

\title{Stokes Parameters as a Minkowskian Four-vector}

\author{D. Han\footnote{electronic mail: han@trmm.gsfc.nasa.gov}}
\address{National Aeronautics and Space Administration, Goddard Space
Flight Center, Code 910.1, Greenbelt, Maryland 20771}

\author{Y. S. Kim\footnote{electronic mail: kim@umdhep.umd.edu}}
\address{Department of Physics, University of Maryland, College Park,
Maryland 20742}

\author{Marilyn E. Noz \footnote{electronic mail:
noz@nucmed.med.nyu.edu}}
\address{Department of Radiology, New York University, New York,
New York 10016}

\maketitle

\begin{abstract}
It is noted that the Jones-matrix formalism for polarization optics
is a six-parameter two-by-two representation of the Lorentz group.
It is shown that the four independent Stokes parameters form a
Minkowskian four-vector, just like the energy-momentum four-vector
in special relativity.  The optical filters are represented by
four-by-four Lorentz-transformation matrices.  This four-by-four
formalism can deal with partial coherence described by the Stokes
parameters.  A four-by-four matrix formulation is given for decoherence
effects on the Stokes parameters, and a possible experiment is
proposed.  It is shown also that this Lorentz-group formalism leads
to optical filters with a symmetry property corresponding to that of
two-dimensional Euclidean transformations.

\end{abstract}

\narrowtext

\section{Introduction}\label{intro}
There are two standard mathematical devices in polarization optics.
One is the Jones-matrix formalism~\cite{jones41}, and the other is the
set of four Stokes parameters~\cite{shur62,hecht70}.  They are
well-established languages in optics.  In our earlier
papers~\cite{hkn96,hkn97josa}, we have shown that the two-by-two
Jones-matrix formalism is basically a two-by-two representation of
the six-parameter Lorentz group, and the Jones vector is like the
two-component spinor.  This formalism allows us to use all the
convenient theorems in the Lorentz group and allows us also to use
optical polarizers as analog computers for the Lorentz group
used in other branches of physics.

It is often more convenient to use the set of four numbers called
the Stokes parameters.  Then it is natural to write down four-by-four
matrices for optical filters.  The purpose of this paper is
to study the parameters and the four-by-four matrices as entities in
the Lorentz group.  We shall show that the Stokes parameters can be
grouped into a Minkowskian four-vector and the four-by-four matrices
are like Lorentz transformation matrices applicable to the
four-dimensional space-time.

Unlike the Jones vector, the Stokes parameters can deal with the
degree of coherence between two orthogonal polarization axes.  In
the language of the two-by-two coherency matrix~\cite{born80},
the lack of coherence appears as a decrease in the magnitude of
the off-diagonal elements.  We shall study this decoherence
effect in detail and propose an experiment to test a Lorentz
effect which is mathematically equivalent to the Thomas effect
in atomic spectra.

For this purpose, we start with the Stokes parameters for a purely
coherent case.  We then study the modification needed to deal with
partially coherent polarized waves.  For the purely coherent case,
it is possible to construct a set of the Stokes parameters from the
Jones vectors.  This process is like constructing a four-vector from
a pair of spinors.   It is also possible to construct four-by-four
filter matrices from the two-by-two Jones matrices.  Thus the
four-by-four matrices applicable to the four Stokes parameters
form another representation of the six-parameter Lorentz group.

It is a simple matter to write down the Stokes parameters for
reduced coherence.  If this reduced coherence is preserved, the
filter matrices are still the four-by-four matrices represented by
the six-parameter Lorentz group.  However, if the optical filters
reduce coherence, this creates an entirely new problem.  In order
to deal with this case, we introduce a four-by-four ``decoherence''
matrix.

This new matrix is not a member of the Lorentz group, and creates
a new matrix algebra.  Fortunately, the four-by-four decoherence
matrix can be reduced to two two-by-two matrices.  One of them
squeezes the first and the fourth Stokes parameters, and the other
two-by-two matrix performs a two-dimensional symplectic transformation
on the second and third Stokes parameters.  This is quite like a
Lorentz transformation in one time-like dimension and two space-like.
Thus, two repeated decoherence matrices can produce a Lorentz effect
similar to the Thomas effect observed in atomic spectra.  This is an
observable effect.

In this paper, we combine various existing mathematical devices
for polarization optics into a single formalism based on the Lorentz
group.  This process leads to some new concepts which can be
observed in laboratories.  The new concepts are decoherence matrices
and new optical filters.

In Sec. \ref{formul}, we construct the two-by-two Jones-matrix
formalism and the Stokes parameters starting from a polarized light
wave.  Section \ref{jones} summarizes our earlier papers on the
Jones-matrix formalism as a representation of the Lorentz group.
In Sec. \ref{trstokes}, we show that the Stokes parameters are
transformed as a Minkowskian four-vector through coherence preserving
filters.

In Sec.~\ref{spinf}, we show how a Minkowskian four-vector is
constructed from two spinors.  It is noted that there are two
different sets of spinors for the case of the Lorentz group, and this
distinction is responsible for the difference between electrons and
positrons.  It is shown how the light wave and its complex conjugate
correspond to the electron and positron respectively.  In
Sec.~\ref{jonsto}, we compare the Jones vectors and the Stokes
parameters, and explain why the four-vector representation of
the Stokes parameters contain more physics than the Jones-matrix
formalism.  It is noted in Sec. \ref{deco} that the Stokes
four-vector can accommodate polarized waves with partial coherence.
We present in this section a matrix formulation of decoherence effects,
and derive a measurable consequence.

In Sec.~\ref{further} we discuss physical implications of the procedure
outlined in this paper, using a concrete physical example.  In
Appendix~A, the Lorentz group is discussed in terms of Lorentz
transformations applicable to the four-dimensional Minkowskian space.
The two-dimensional Euclidean group is used in our discussion of
possible new optical filters.  We explain this group in terms of
translations and rotations on a flat plane in Appendix B.

\section{Formulation of the Problem}\label{formul}
In studying polarized light propagating
along the $z$ direction, the traditional approach is to consider the $x$
and $y$ components of the electric fields.  Their amplitude ratio and the
phase difference determine the degree of polarization.  Thus, we can
change the polarization either by adjusting the amplitudes, by changing
the relative phases, or both.  For convenience, we call the optical
device which changes amplitudes an ``attenuator'' and the device which
changes the relative phase a ``phase shifter.''

Let us write these electric fields as
\begin{equation}\label{expo1}
\pmatrix{E_{x} \cr E_{y}} =
\pmatrix{A \exp{\left\{i(kz - \omega t + \phi_{1})\right\}}  \cr
B \exp{\left\{i(kz - \omega t + \phi_{2})\right\}}} .
\end{equation}
where $A$ and $B$ are the amplitudes which are real and positive
numbers, and $\phi_{1}$ and $\phi_{2}$ are the phases of the $x$ and
$y$ components respectively.  This column matrix is called the Jones
vector.  The content of polarization is determined by the ratio:
\begin{equation}
{E_{y}\over E_{x}} = \left({B\over A}\right) e^{i(\phi_{2} - \phi_{1})} .
\end{equation}
which can be written as one complex number:
\begin{equation}\label{ratio}
w = r e^{i\phi}
\end{equation}
with
$$
r = {B \over A} , \qquad \phi = \phi_{2} - \phi_{1} .
$$
The degree of polarization is measured by these two real numbers, which
are the amplitude ratio and the phase difference respectively.
The transformation takes place when the light
beam goes through an optical filter whose transmission properties are
not isotropic.

In dealing with light waves, we have to realize that the intensity
is the quantity we measure.  Then there arises the question of
coherence and time average.  We are thus led to consider the following
parameters.
\begin{eqnarray}\label{sii}
S_{11} &=& <E_{x}^{*}E_{x}>  , \qquad
S_{22} = <E_{y}^{*}E_{y}> , \cr
S_{12} &=& <E_{x}^{*}E_{y}> ,  \qquad
S_{21} = <E_{y}^{*}E_{x}> .
\end{eqnarray}
Then, we are naturally invited to write down the two-by-two matrix:
\begin{equation}\label{cohm1}
C = \pmatrix{<E^{*}_{x}E_{x}> & <E^{*}_{y} E_{x}> \cr
<E^{*}_{x} E_{y}> & <E^{*}_{y} E_{y}>} ,
\end{equation}
where $<E^{*}_{i}E_{j}>$ is the time average of $E^{*}_{i}E_{j}$.
The above form is called the coherency matrix~\cite{born80}.

It is sometimes more convenient to use the following combinations of
parameters.
\begin{eqnarray}\label{stokes}
S_{0} &=& S_{11} + S_{22}, \cr
S_{1} &=& S_{11} - S_{22}, \cr
S_{2} &=& S_{12} + S_{21}, \cr
S_{3} &=& -i\left(S_{12} - S_{21}\right),
\end{eqnarray}
These four parameters are called the Stokes parameters in the
literature~\cite{shur62,hecht70}.

The purpose of the present paper is to show that these four
parameters form a Minkowskian four-vector when the light wave goes
through optical filters.
Once the four Stokes parameters are introduced, it is quite natural
to construct a four-dimensional vector and transformation matrices
applicable to this vector.  These four-by-four matrices are called
Mueller matrices~\cite{shur62}.  In this paper, we shall therefore
show that the Mueller matrices are like four-by-four Lorentz
transformation matrices.

G. G. Stokes was an active researcher during the 19th
Century~\cite{shur62}.  The mathematics of the Stokes parameters
is a very old science, and there are many interesting mathematical
devices.  For instance, we can define a three-dimensional Cartesian
coordinate system spanned by $S_{1}$, $S_{2}$, $S_{3}$.  We can then
consider a sphere whose maximum radius is $S_{0}$.  This is called
the Poincar\'e sphere and is the standard geometrical language for the
Stokes parameters~\cite{born80}.  We shall use the geometry of this
sphere in Secs.~\ref{jonsto} and \ref{deco} of this paper.

\section{Jones-matrix Formalism}\label{jones}
We presented in our earlier papers~\cite{hkn96}, the Jones-matrix
formalism as a representation of the Lorentz group.  We used there
the concept of squeeze transformations to describe attenuation
filters.  Let us summarize in this section the Lorentz-group
content of this formalism which will be useful for discussing
the Stokes parameters.

There are two transverse directions which are perpendicular to each
other.  The absorption coefficient in one transverse direction could
be different from the coefficient along the other direction.  Thus,
there is the ``polarization'' coordinate in which the absorption can
be described by
\begin{equation}\label{atten}
\pmatrix{e^{-\eta_{1}} & 0 \cr 0 & e^{-\eta_{2}}} =
e^{-(\eta_{1} + \eta_{2})/2} \pmatrix{e^{\eta/2} & 0 \cr
0 & e^{-\eta/2}}
\end{equation}
with $\eta = \eta_{2} - \eta_{1}$ .  This attenuation matrix tells us
that the electric fields are attenuated at two different rates.  The
exponential factor $e^{-(\eta_{1} + \eta_{2})/2}$ reduces both
components at the same rate and does not affect the degree of
polarization.  The effect of polarization is solely determined by the
attenuation matrix
\begin{equation}\label{at1}
A(0, \eta) = \pmatrix{e^{\eta/2} & 0 \cr 0 & e^{-\eta/2}} .
\end{equation}
This type of mathematical operation is quite familiar to us from
squeezed states of light and from Lorentz boosts of spinors.
We call the above matrix the ``attenuator.''

Another basic element is the optical filter with different values of
the index of refraction along the two orthogonal directions.  The
effect of this filter can be written as
\begin{equation}\label{phase}
\pmatrix{e^{i\delta_{1}} & 0 \cr 0 & e^{i\delta_{2}}}
= e^{-i(\delta_{1} + \delta_{2})/2}
\pmatrix{e^{-i\delta/2} & 0 \cr 0 & e^{i\delta/2}} ,
\end{equation}
with $\delta = \delta_{2} - \delta_{1}$ .
In measurement processes, the overall phase factor
$e^{-i(\delta_{1} + \delta_{2})/2}$
cannot be detected, and can therefore be deleted.  The polarization
effect of the filter is solely determined by the matrix
\begin{equation}\label{shif1}
P(0, \delta) = \pmatrix{e^{-i\delta/2} & 0 \cr 0 & e^{i\delta/2}} .
\end{equation}
This form was noted as one of the basic components of the Jones-matrix
formalism in Sec.~\ref{formul}.  This phase-shifter matrix appears
like a rotation matrix around the $z$ axis in the theory of rotation
groups, but it plays a different role in this paper.  We shall
hereafter call this matrix a phase shifter.

The polarization axes are not always the $x$ and $y$ axes.  For this
reason, we need the rotation matrix
\begin{equation}\label{rot1}
R(\theta) = \pmatrix{\cos(\theta/2) & -\sin(\theta/2)
\cr \sin(\theta/2) & \cos(\theta/2)} .
\end{equation}
If the polarization coordinate is rotated by an angle $\theta/2$, the
attenuator and phase-shifter of Eq.(\ref{at1}) and Eq.(\ref{shif1})
respectively become
\widetext
\begin{eqnarray}\label{at2}
A(\theta, \eta) &=& R(\theta) A(0, \eta) R(-\theta) \nonumber  \\[2mm]
&=& \pmatrix{e^{\eta/2}\cos^{2}(\theta/2) + e^{-\eta/2}\sin^{2}(\theta/2) &
(e^{\eta/2} - e^{-\eta/2})\cos(\theta/2) \sin(\theta/2) \cr
(e^{\eta/2} - e^{-\eta/2})\cos(\theta/2) \sin(\theta/2)
& e^{-\eta/2}\cos^{2}(\theta/2) + e^{\eta/2}\sin^{2}(\theta/2)} ,
\end{eqnarray}
and
\begin{eqnarray}\label{shif2}
P(\theta, \delta) &=& R(\theta)
P(0, \delta) R(-\theta) \nonumber \\[2mm]
& = & \pmatrix{e^{-i\delta/2}\cos^{2}(\theta/2) +
e^{i\delta/2}\sin^{2}(\theta/2) &
(e^{-i\delta/2} - e^{i\delta/2})\cos(\theta/2) \sin(\theta/2) \cr
(e^{-i\delta/2} - e^{i\delta/2})\cos(\theta/2) \sin(\theta/2)
& e^{i\delta/2}\cos^{2}(\theta/2) + e^{-i\delta/2}\sin^{2}(\theta/2)} .
\end{eqnarray}
\narrowtext

The Jones-matrix formalism consists of repeated applications of the
three basic operations.  In order to study this more systematically,
let us use the generators of these transformations.  First, the
rotation matrix of Eq.(\ref{rot1}) takes the form:
\begin{equation}
R(\theta) = \exp{\left(-i\theta J_{2} \right)} ,
\end{equation}
with
\begin{equation}\label{j2}
J_{2} = {1 \over 2} \pmatrix{0 & -i \cr i & 0} .
\end{equation}
The attenuation operator of Eq.(\ref{at1}) can also be written in the
exponential form:
\begin{equation}
A(0,\eta) = \exp{\left(-i\eta K_{3} \right)} ,
\end{equation}
with $$ K_{3} = {i \over 2} \pmatrix{1 & 0 \cr 0 & -1} . $$

If we take the commutation relation between the above generators, we
end up with another generator:
\begin{equation}
[J_{2}, K_{3}] = iK_{1} ,
\end{equation}
with $$K_{1} = {i\over 2} \pmatrix{0 & 1 \cr 1 & 0} . $$
We are then led to write the commutation relations:
\begin{equation}
[K_{3}, K_{1}] = - iJ_{2} , \qquad
[K_{1}, J_{2}] = iK_{3} .
\end{equation}
These three generators indeed form a closed set of commutation
relations.  The three-parameter groups generated by this set of
commutation relations are $O(2,1)$, $Sp(2,r)$, and $SU(1,1)$.  The
group $SU(1,1)$ has been extensively discussed in the literature in
connection with squeezed states of light~\cite{knp91}.  The group
$O(2,1)$ is the Lorentz group applicable to one time-like dimension
and two space-like dimensions.  The group $Sp(2,r)$ is the basic
symmetry group for the Wigner function for one-mode squeezed
states~\cite{knp91,hkn88}.  The relevance of this group in
polarization optics has also been noted in the
literature~\cite{hkn96,chiao88,kitano89}.

The phase-shifter of Eq.(\ref{shif1}) can be written as
\begin{equation}
P(0,\delta) = \exp{\left(-i\delta J_{3} \right)} ,
\end{equation}
with
\begin{equation}\label{j3}
J_{3} = {1 \over 2} \pmatrix{1 & 0 \cr 0 & -1} .
\end{equation}
If we take a commutation relation of this operator with $J_{2}$,
\begin{equation}
[J_{3}, J_{2}] = iJ_{1} ,
\end{equation}
with $$J_{1} = {1\over 2} \pmatrix{0 & 1 \cr 1 & 0} . $$
These three generators satisfy the commutation relations
\begin{equation}\label{commjjj}
[J_{i}, J_{j}] = i\epsilon_{ijk} J_{k} .
\end{equation}
This set of commutations relations is very familiar to us.  They
generate the three-dimensional rotation group and the $SU(2)$ group
governing rotational symmetry of spin-1/2 particles.  These are also
three-parameter groups.

If we apply both attenuators and phase-shifters in random order in
laboratories, the corresponding mathematics is to combine the two
three-parameter groups by mixing up their commutation relations.  The
resulting closed set of commutation relations consists of~\cite{hkn96}
\begin{equation}
[J_{i}, K_{j}] = i\epsilon_{ijk} K_{k} , \qquad
[K_{i}, J_{j}] = -i\epsilon_{ijk} J_{k} ,
\end{equation}
in addition to the set given in Eq.(\ref{commjjj}).

These are the six generators for the two-by-two representation of the
Lorentz group which is often called $SL(2,c)$.  This group is the
standard language in elementary particle physics dealing with spin-1/2
particles.  In optics, the Lorentz group gained its prominence
recently in connection with squeezed states of light~\cite{knp91}.
It is quite natural for us to study the two-by-two Jones matrices
and four-by-four Mueller matrices within the framework of the
Lorentz group.

\section{Transformation Properties of the Stokes
Parameters}\label{trstokes}
In order to study the effect of each filter, let us note that the
effect of the attenuator of Eq.(\ref{at1}) on the incoming light of
Eq.(\ref{expo1}) is
\begin{equation}\label{effect1}
\pmatrix{e^{\eta/2} E_{x} \cr e^{-\eta/2} E_{y}} .
\end{equation}
The effect of the phase shifter of Eq.(\ref{shif1})
on the incoming light of Eq.(\ref{expo1}) is
\begin{equation}\label{effect2}
\pmatrix{e^{-i\delta/2} E_{x} \cr e^{i\delta/2} E_{y} } .
\end{equation}
The effect of the rotation matrix of Eq.(\ref{rot1}) on the incoming
light wave is
\begin{equation}\label{effect3}
\pmatrix{(\cos(\theta/2)) E_{x} - (\sin(\theta/2)) E_{y}
\cr (\sin(\theta/2)) E_{x} + (\cos(\theta/2)) E_{y}} .
\end{equation}

Under the rotation of Eq.(\ref{rot1}) which transforms the incoming
wave of Eq.(\ref{expo1}) to Eq.(\ref{effect3}), $S_{11}$ and $S_{22}$
become
\begin{equation}
{\cos\theta\over2} (S_{11} - S_{22})
- {\sin\theta\over 2} (S_{12} + S_{21})
+ {1\over2} (S_{11} + S_{22}) ,
\end{equation}
and
\begin{equation}
- {\cos\theta\over2} (S_{11} - S_{22})
+ {\sin\theta\over 2} (S_{12} + S_{21})
+ {1\over2} (S_{11} + S_{22}) ,
\end{equation}
respectively.  It is clear from these expressions that
$(S_{11} + S_{22})$ is invariant under this rotation.  As for
$(S_{11} - S_{22})$, the rotation leads to
\begin{equation}
\cos\theta (S_{11} - S_{22}) - \sin\theta (S_{12} + S_{21}) .
\end{equation}
Under the same rotation, $S_{12}$ and $S_{21}$ become
\begin{equation}
{\sin\theta \over 2} (S_{11} - S_{22}) +
{\cos\theta \over 2} (S_{12} + S_{21}) + {1\over 2}(S_{12} - S_{21}) ,
\end{equation}
and
\begin{equation}
{\sin\theta \over 2} (S_{11} - S_{22}) +
{\cos\theta \over 2} (S_{12} + S_{21}) - {1\over 2}(S_{12} - S_{21}) ,
\end{equation}
respectively.  Thus $(S_{12} - S_{21})$ remains invariant, while
$(S_{12} + S_{21})$ becomes
\begin{equation}
\sin\theta (S_{11} - S_{22}) + \cos\theta (S_{12} + S_{21}) .
\end{equation}
We can thus write the effect of the rotation as
\begin{equation}
\pmatrix{\cos\theta & -\sin\theta \cr \sin\theta & \cos\theta}
\pmatrix{S_{11} - S_{22} \cr S_{12} + S_{21}} .
\end{equation}

Under the phase-shift transformation of Eq.(\ref{shif1}) which
leads to Eq.(\ref{effect2}), $S_{11}$ and $S_{22}$ remain invariant,
while $S_{12}$ and $S_{21}$ become $e^{i\delta}S_{12}$ and
$e^{-i\delta}S_{21}$.  The result is
\begin{equation}
\pmatrix{\cos\delta & -\sin\delta \cr \sin\delta & \cos\delta}
\pmatrix{S_{12} + S_{21} \cr -i(S_{12} - S_{21})} .
\end{equation}
Under the squeeze transformation of Eq.(\ref{at1}) which leads to
Eq.(\ref{effect1}), $S_{11}$ and $S_{22}$ become $e^{\eta}S_{11}$ and
$e^{-\eta}S_{22}$ respectively, while $S_{12}$ and $S_{21}$ remain
unchanged.  This can be translated into
\begin{equation}
\pmatrix{\cosh\eta & \sinh\eta \cr \sinh\eta & \cosh\eta}
\pmatrix{S_{11} + S_{22} \cr S_{11} - S_{22}} .
\end{equation}

Thus in terms of the Stokes parameters given in Eq.(\ref{stokes}),
we can write the above three transformations as
\begin{equation}\label{tr1}
\pmatrix{1 & 0 & 0 & 0 \cr 0 & \cos\theta & -\sin\theta & 0 \cr
0 & \sin\theta & \cos\theta & 0 \cr 0 & 0 & 0 & 1}
\pmatrix{S_{0} \cr S_{1} \cr S_{2} \cr S_{3}} ,
\end{equation}
\begin{equation}\label{tr2}
\pmatrix{1 & 0 & 0 & 0 \cr 0 & 1 & 0 & 0 \cr
0 & 0 & \cos\delta & -\sin\delta \cr
0 & 0 & \sin\delta & \cos\delta}
\pmatrix{S_{0} \cr S_{1} \cr S_{2} \cr S_{3}} ,
\end{equation}
and
\begin{equation}\label{tr3}
\pmatrix{\cosh\eta & \sinh\eta & 0 & 0
\cr \sinh\eta & \cosh\eta & 0 & 0 \cr
0 & 0 & 1 & 0 \cr 0 & 0 & 0 & 1}
\pmatrix{S_{0} \cr S_{1} \cr S_{2} \cr S_{3}} .
\end{equation}

The above three matrices are generated by $J_{3}$, $J_{1}$ and
$K_{1}$ respectively, where
\begin{equation}\label{jj31}
J_{3} = \pmatrix{0 & 0 & 0 & 0 \cr 0 & 0 & -i & 0 \cr
0 & i & 0 & 0 \cr 0 & 0 & 0 & 0}, \quad
J_{1} = \pmatrix{0 & 0 & 0 & 0 \cr 0 & 0 & 0 & 0 \cr
0 & 0 & 0 & -i \cr 0 & 0 & i & 0},
\end{equation}
and
\begin{equation}\label{kk1}
K_{1} = \pmatrix{0 & i & 0 & 0 \cr i & 0 & 0 & 0 \cr
0 & 0 & 0 & 0 \cr 0 & 0 & 0 & 0} .
\end{equation}
The commutation relation between $J_{3}$ and $J_{1}$ leads to another
generator $J_{2}$, with
\begin{equation}\label{jj2}
J_{2} = \pmatrix{0 & 0 & 0 & 0 \cr 0 & 0 & 0 & i \cr
0 & 0 & 0 & 0 \cr 0 & -i & 0 & 0} .
\end{equation}
These three matrices generate the three-dimensional rotation group
with the closed set of commutation relations.
\begin{equation}\label{comm1}
\left[J_{i}, J_{j}\right] = i \epsilon_{ijk} J_{k} .
\end{equation}
If we take commutations relations of $K_{1}$ with these rotation
generators, we end up with two additional generators $K_{2}$ and
$K_{3}$, where
\begin{equation}\label{kk23}
K_{2} = \pmatrix{0 & 0 & i & 0 \cr 0 & 0 & 0 & 0 \cr
i & 0 & 0 & 0 \cr 0 & 0 & 0 & 0}, \quad
K_{3} = \pmatrix{0 & 0 & 0 & i \cr 0 & 0 & 0 & 0 \cr
0 & 0 & 0 & 0 \cr i & 0 & 0 & 0} .
\end{equation}
These three $K_{i}$ generators satisfy the commutation relations
\begin{equation}\label{comm2}
\left[K_{i}, K_{j}\right] = -i \epsilon_{ijk} J_{k} .
\end{equation}
The right-hand side of the above expression is not $K_{i}$ but $J_{i}$.
Thus, in order to get a closed set of commutators, we have to take
commutation relations between $J_{i}$ and $K_{i}$.  The result is
\begin{equation}\label{comm3}
\left[J_{i}, K_{j}\right] = i \epsilon_{ijk} K_{k} .
\end{equation}
Thus the six matrices consisting of three $J$ and three $K$ matrices
form a closed set of a group of transformations applicable to
four-dimensional space.

If the above transformation group is applied to the four-dimensional
space-time coordinate $(t, z, x, y)$, it becomes the group of
Lorentz transformations in the Minkowskian space.  In deriving the
above three sets of commutation relations, we have not used any of
the principles of special relativity.  The commutation relations are
derived strictly from the properties of the optical filters.  The
Stokes parameters have nothing to do with special relativity.  Yet,
it is remarkable that they can be formulated in terms of the
mathematics of Lorentz transformations.

\section{Spinors and Four-vectors in the Lorentz Group}\label{spinf}
We are now confronted with the question of why the Stokes parameters
have to behave like a Minkowskian four-vector.  For this purpose,
let us go back to Sec. \ref{jones} and consider repeated applications
of the three basic operations.  We shall see first whether the
two-by-two matrix algebra of Sec. \ref{jones} can be represented
as a representation of the six-parameter Lorentz group.  We shall then
investigate whether the four-by-four matrices of Sec.~\ref{trstokes}
can be systematically obtained from the Jones matrices.

For this purpose, let us start with the three Pauli spin matrices
of the form
\begin{equation}
\sigma_{1} = \pmatrix{0 & 1 \cr 1 & 0} , \quad
\sigma_{2} = \pmatrix{0 & -i \cr i & 0} , \quad
\sigma_{3} = \pmatrix{1 & 0 \cr 0 & -1} .
\end{equation}
Then, the rotation generators
\begin{equation}
J_{i} = {1 \over 2} \sigma_{i}
\end{equation}
satisfy the closed set of commutation relations of Eq.(\ref{comm1}).

We can also construct three boost generators
\begin{equation}
K_{i} = {i \over 2} \sigma_{i} ,
\end{equation}
which satisfy the commutation relations given in Eq.(\ref{comm2}).
The $K_{i}$ matrices alone do not form a closed set of commutation
relations, their commutation relations with the rotation generators
are given in Eq.(\ref{comm3}).

The six matrices $J_{i}$ and $K_{i}$ form a closed set of commutation
relations, and they are like the generators of the Lorentz group
applicable to the (3 + 1)-dimensional Minkowski space.  The group
generated by the above six matrices is called $SL(2,c)$ consisting of
all two-by-two complex matrices with unit determinant.

In order to construct four-vectors, we need two different spinor
representations of the Lorentz group.  Let us go to the commutation
relations for the generators given in Eqs.(\ref{comm1}), (\ref{comm2})
and (\ref{comm3}).  These commutators are not invariant under the
sign change of the rotation generators $J_{i}$, but are invariant
under the sign change of the squeeze generators $K_{i}$.  Thus, to
each spinor representation, there is another representation with the
squeeze generators with opposite sign.  This allows us to construct
another representation with the generators:
\begin{equation}
\dot{J}_{i} = {1 \over 2} \sigma_{i}, \qquad
\dot{K}_{i} = {-i \over 2} \sigma_{i} .
\end{equation}
We call this representation the ``dotted'' representation.

There are therefore two different sets of Lorentz transformation
matrices.  If we write the most general form of the transformation
matrix using undotted generators, it takes the form
\begin{equation}\label{elundot}
L = \exp\left\{-{i\over 2} \sum_{i=1}^{3}\left(\theta_{i}\sigma_{i} +
i\eta_{i}\sigma_{i}\right) \right\} .
\end{equation}
Then the transformation matrix in the dotted representation becomes
\begin{equation}\label{eldot}
\dot{L} = \exp\left\{-{i\over 2} \sum_{i=1}^{3}\left(\theta_{i}\sigma_{i}
- i\eta_{i}\sigma_{i}\right)\right\} .
\end{equation}
In both of the above matrices, the Hermitian conjugation changes the
direction of rotation.  However, it does not change the direction of
boosts.  We can achieve this only by interchanging $L$ to $\dot{L}$,
and we shall call this the ``dot'' conjugation.

Likewise, there are two different sets of spinors.  Let us use $u$ and
$v$ for the up and down spinors for ``undotted'' representation.  Then
$\dot{u}$ and $\dot{v}$ are for the dotted representation.  The
four-vectors are then constructed as~\cite{hks86}
\begin{eqnarray}
u\dot{u} &=& - (x - iy), \quad v\dot{v} = (x + iy), \cr
u\dot{v} &=& (t + z), \quad v\dot{u} = -(t - z) .
\end{eqnarray}

The relation between the $SL(2,c)$ spinors and the four-vectors has been
discussed in the literature~\cite{hks86,knp86,gelf63}.
It is possible to construct the four-vector with the four
$SL(2,c)$ spinors~\cite{knp86,hks86jm}.  Indeed,
\begin{eqnarray}
-u\dot{u} &=& (1, i, 0, 0) ,\hspace{10mm}
v\dot{v} = (1, -i, 0, 0) , \cr
u\dot{v} &=& (0, 0, 1, 1) , \hspace{10mm}
v\dot{v}_{+} = (0, 0, 1, -1) .
\end{eqnarray}

It is possible to construct a six-component Maxwell tensor by making
combinations of two undotted and dotted spinors~\cite{knp86}.  For
massless particles, the only gauge-invariant components are $uu$ and
$\dot{v}\dot{v}$~\cite{wein64}.  They correspond to the photons in the
Maxwell tensor representation with positive and negative helicities
respectively.  It is also possible to construct Maxwell-tensor fields
only for a massive particle, and obtain massless Maxwell fields by
group contraction~\cite{bask95}.

\begin{equation}\label{dotmat}
C = \pmatrix{u \dot{v} & -u\dot{u} \cr v\dot{v} & -v\dot{u}}
   = \pmatrix{u \cr v} \pmatrix {\dot{v} & -\dot{u}} ,
\end{equation}
where $u$ and $\dot{u}$ are one if the spin is up, and are zero if the
spin is down, while $v$ and $\dot{v}$ are zero and one for the spin-up
and spin-down cases.
The transformation matrix applicable to the column vector in the above
expression is the two-by-two matrix given in Eq.(\ref{elundot}).  What
is then the transformation matrix applicable to the row vector
$(\dot{v},~-\dot{u})$ from the right-hand side?  It is the transpose
of the matrix applicable to the column vector $(\dot{v},~-\dot{u})$.
We can obtain this column vector from
\begin{equation}\label{dotcol}
 \pmatrix {\dot{v} \cr -\dot{u}} ,
\end{equation}
by applying to it the matrix
\begin{equation}
g = -i\sigma_{2} = \pmatrix{0 & -1 \cr 1 & 0} .
\end{equation}
This matrix also has the property
\begin{equation}
g \sigma_{i} g^{-1} = -\left(\sigma_{i}\right)^{T} ,
\end{equation}
where the superscript $T$ means the transpose of the matrix.  The
transformation matrix applicable to the column vector of Eq.(\ref{dotcol})
is $\dot{L}$ of Eq.(\ref{eldot}).  Thus the matrix applicable to the row
vector $(\dot{v},~-\dot{u})$ in Eq.(\ref{dotmat}) is
\begin{equation}
\left\{g^{-1} \dot{L} g\right\}^{T} = g^{-1} \dot{L}^{T} g .
\end{equation}
This is precisely the Hermitian conjugate of $L$.

Let us now consider its transformation properties.
The matrix of Eq.(\ref{cohm1}) is like
\begin{equation}\label{matC}
C = \pmatrix{t + z & x - iy \cr x + iy & t - z} ,
\end{equation}
where the set of variables $(x, y, z, t)$ is transformed like a four-vector
under Lorentz transformations.  Furthermore, it is known that the Lorentz
transformation of this four-vector is achieved through the formula
\begin{equation}\label{Ldag}
C' = L C L^{\dagger} ,
\end{equation}
where the transformation matrix $L$ is that of Eq.(\ref{elundot}).
The construction of four-vectors from the two-component spinors is not
a trivial task~\cite{hks86,bask95}.  The two-by-two representation of
Eq.(\ref{matC}) requires one more step of complication.

We are not the first ones to suspect that the coherency matrix behaves
like a four-vector.  This was done by Barakat in 1963~\cite{barakat63}.
However, Sec. \ref{spinf} shows that we need two different two-by-two
representations of the Lorentz group to establish the connection
between the Jones vectors and the Stokes parameters in a covariant
manner.  The two different spinors are the column vectors of
$(u, v)$ and $(\dot{u}, \dot{v})$.

This paper allows us to associate these spinors as
\begin{equation}\label{uv}
\pmatrix{u \cr v} = \pmatrix{E_{x} \cr E_{y}} ,
\end{equation}
and
\begin{equation}\label{uvdot}
\pmatrix{\dot{u} \cr \dot{v}} = \pmatrix{-E_{y}^{*} \cr E_{x}^{*}} .
\end{equation}
The symmetry between the dotted and undotted representations is
responsible for the electron-positron symmetry in the Dirac
equation~\cite{knp86}.  It is interesting to note that this symmetry
is applicable also to the polarization vectors of Eq.(\ref{uv}) and
Eq.(\ref{uvdot}).

\section{Jones Spinors and Stokes Vectors}\label{jonsto}
The Jones vector is a two-component vector in the conventional
formalism.  Since, however, it is like a spinor in the Lorentz
group, we call it hereafter the Jones spinor.  The Jones spinor
and the Stokes four-vector are two different representations of
the same Lorentz group.  Why do we construct two different
representations?  The difference is in physics.

Since the four-vector contains more elements than the two-component
spinor, the Stokes vector should give more information than the Jones
spinor.  This is translated into the invariance properties of Stokes
parameters.  As the four-scalar $(t^{2} - z^{2}- x^{2} - y^{z})$ is
invariant under Lorentz transformations, the quantity
\begin{equation}
S^{2} = S_{0}^{2} - S_{1}^{2} - S_{2}^{2} - S_{3}^{2}
\end{equation}
remains invariant under filtering processes discussed in this paper.
We shall hereafter call this quantity ``Stokes scalar.''  If the
Stokes Scalar is zero, the system is completely coherent.  This scalar
quantity is positive if the system is partially coherent.

Indeed, this degree of coherence is what the Stokes vector can tell
while the Jones spinor cannot.  If the filter system leaves the Stokes
scalar invariant, it is a coherence-preserving system.  This quantity
is not preserved if the filters cause random variations of phases.
The best way to describe this degree of coherence is to construct
a Poincar\'e sphere in the three-dimensional space of $S_{1}, S_{2}$
and $S_{3}$.  The radius of this sphere is
\begin{equation}
R = \left(S_{1}^{2} + S_{2}^{2} + S_{3}^{2}\right)^{1/2} .
\end{equation}
Then the ratio $R/S_{0}$ gives the degree of coherence.

This radius takes the maximum value $S_{0}$ when the system is
completely coherent, and it takes the minimum value of $S_{1}$ when
the system is completely incoherent.  In this minimum case, both
$S_{2}$ and $S_{3}$ vanish.  The question then is whether there is
a four-by-four matrix which reduces these two components.  If so,
how can this matrix be augmented to the set of transformation
matrices discussed in Sec. \ref{trstokes}?

\section{Decoherence Matrices}\label{deco}
Let us go back to the four-by-four representation of Sec. \ref{trstokes}.
For the Stokes four-vector, we can translate the two-by-two attenuator
Eq.(\ref{at1}) applicable to the Jones vector into the four-by-four
matrix
\begin{equation}\label{at4}
A(0, \eta) = \pmatrix{\cosh\eta & \sinh\eta & 0 & 0 \cr
\sinh\eta & \cosh\eta & 0 & 0 \cr 0 & 0 & 1 & 0 \cr 0 & 0 & 0 & 1} .
\end{equation}
Likewise, the phase-shifter of Eq.(\ref{shif1}) is translated into
\begin{equation}\label{shif4}
P(0, \delta) = \pmatrix{1 & 0 & 0 & 0 \cr 0 & 1  & 0 & 0 \cr
0 & 0 & \cos\delta & -\sin\delta \cr 0 & 0 & \sin\delta & \cos\delta} .
\end{equation}
The rotation matrix of Eq.(\ref{rot1}) becomes
\begin{equation}\label{rot4}
R(\theta) = \pmatrix{1 & 0 & 0 & 0 \cr 0 & \cos\theta & -\sin\theta &
0 \cr 0 & \sin\theta & \cos\theta & 0 \cr 0 & 0 & 0 & 1} .
\end{equation}

If the two transverse components lose coherence, the time-averaged
values $S_{12}$ and $S_{21}$ become smaller.  We can therefore use
the matrix
\begin{equation}
\pmatrix{1 & 0 & 0 & 0 \cr 0 & 1 & 0 & 0 \cr
0 & 0 & e^{-2\lambda} & 0 \cr 0 & 0 & 0 & e^{-2\lambda}} ,
\end{equation}
which can also be written as
\begin{equation}\label{decoh1}
e^{-\lambda} \pmatrix{e^{\lambda} & 0 & 0 & 0 \cr
0 & e^{\lambda} & 0 & 0 \cr 0 & 0 & e^{-\lambda} & 0 \cr
0 & 0 & 0 & e^{-\lambda}} ,
\end{equation}
where $e^{-\lambda}$ is the overall decoherence factor.  For convenience,
we define the decoherence matrix as
\begin{equation}
D(\lambda) = \pmatrix{e^{\lambda} & 0 & 0 & 0 \cr
0 & e^{\lambda} & 0 & 0 \cr 0 & 0 & e^{-\lambda} & 0 \cr
0 & 0 & 0 & e^{-\lambda}} ,
\end{equation}
which is generated by
\begin{equation}
Q_{3} = \pmatrix{i & 0 & 0 & 0 \cr 0 & i & 0 & 0
\cr 0 & 0 & -i & 0 \cr 0 & 0 & 0 & -i} .
\end{equation}
The introduction of the above matrix into the existing set of six
generators of the Lorentz group leads to the fifteen parameter group
of $O(3,3)$ or $SL(4,r)$~\cite{hkn95jm}, and this is beyond the scope
of the present paper.  This however does not prevent us from looking for
an interesting subgroup which will play the key role in accommodating
the decoherence matrix into the four-by-four matrix formalism for the
Stokes parameters.

It is interesting to see that this decoherence matrix commutes with
the attenuator of Eq.(\ref{at4}) and the phase-shifter of
Eq.(\ref{shif4}), but it does not commute with the rotator of
Eq.(\ref{rot4}).  Thus, the complication is reduced to the
non-commuting algebra of this rotation matrix and the decoherence
matrix.  As is given in Eq.(\ref{jj31}), the generator of $R(\theta)$
of Eq.(\ref{rot4}) takes the form
\begin{equation}
J_{3} = \pmatrix{0 & 0 & 0 & 0 \cr 0 & 0 & -i & 0 \cr
0 & i & 0 & 0 \cr 0 & 0 & 0 & 0} .
\end{equation}
If we take the commutator of this matrix with the generator of $Q_{3}$,
\begin{equation}\label{w3}
[J_{3}, Q_{3}] = 2i W_{3} ,
\end{equation}
with
\begin{equation}
W_{3} = \pmatrix{0 & 0 & 0 & 0 \cr 0 & 0 & i & 0 \cr
0 & i & 0 & 0 \cr 0 & 0 & 0 & 0} .
\end{equation}

In order to see the physics of these matrices, let us go to the
Poincar\'e sphere of this system.  In the three-dimensional space
with the three Cartesian coordinate variable $S_{1}, S_{2}$ and
$S_{3}$, rotations around the $S_{3}$ axis generated by $J_{3}$ do
not change the first and the last components of the Stokes
four-vector $(S_{0}, S_{1}, S_{2}, S_{3})$.  We can thus divide this
four-component vector into two two-component vectors:
\begin{equation}
S_{A} = \pmatrix{S_{0} \cr S_{3}} ,  \qquad
S_{B} = \pmatrix{S_{1} \cr S_{2}} .
\end{equation}
The effect of the decoherence matrix on $S_{A}$ will be
\begin{equation}
\pmatrix{e^{\lambda} & 0 \cr 0 & e^{-\lambda}} \pmatrix{S_{0} \cr S_{3}} .
\end{equation}
This is a squeeze transformation not affected by rotations around the
$S_{3}$ axis.  If we take into account the overall factor mentioned
after Eq.(\ref{decoh1}), the effect of decoherence on $S_{A}$ is
\begin{equation}
\pmatrix{1 & 0 \cr 0 & e^{-2\lambda}} \pmatrix{S_{0} \cr S_{3}} ,
\end{equation}
and this expression is invariant under rotations generated by $J_{3}$.
Under repeated applications, the matrix algebra is simply
\begin{equation}
\pmatrix{1 & 0 \cr 0 & e^{-2\rho}}
\pmatrix{1 & 0 \cr 0 & e^{-2\lambda}}
\pmatrix{S_{0} \cr S_{3}} .
\end{equation}

The story is quite different for $S_{B}$.  Here we are dealing with
rotations and squeeze transformations in the two-dimensional space
of $S_{1}$ and $S_{2}$.  In order to take advantage of the mathematics
of squeezed states~\cite{knp91}, let us write $Q_{3}, W_{3}, J_{3}$ as
\begin{equation}
B_{1} = {1 \over 2} Q_{3}, \quad
B_{2} = {1 \over 2} W_{3}, \quad
L_{3} = {1 \over 2} J_{3} .
\end{equation}
These matrices form the closed set of commutation relations
\begin{equation}
[B_{1}, B_{2}] = -i L_{3} , \quad [B_{2}, L_{3}] = i B_{3} , \quad
[L_{3}, B_{1}] = i B_{2} ,
\end{equation}
which are very familiar to us from the squeezed state of light.  They
generate the group $Sp(2)$ or $SU(1,1)$.  This is by now a standard
mathematical tool in optics.  The algebraic property of this group
is the same as the group of Lorentz transformations in two space-like
dimensions and one time-like dimension.  This group is routinely
called $O(2,1)$ in the literature, and the generators satisfy the same
set of commutation relations as the above set for $Sp(2)$~\cite{knp91}.

Thus the matrix algebra applicable to the two-component vector $S_{B}$
is the same as that for the $Sp(2)$ squeezed states and/or the
$(2~+~1)$-dimensional Lorentz group.  The decoherence along the
$S_{1}$ direction is
\begin{equation}\label{decoh2}
\pmatrix{e^{\lambda} & 0 \cr 0 & e^{-\lambda}} .
\end{equation}
The decoherence transformation along the direction which makes an
angle $\theta$ with the $S_{1}$ axis is
\widetext
\begin{equation}\label{decoh3}
\pmatrix{\cosh\rho + (\sinh\rho)\cos(2\theta) &
(\sinh\rho)\sin(2\theta) \cr (\sinh\rho)\sin(2\theta) &
\cosh\rho - (\sinh\rho)\cos(2\theta) }.
\end{equation}
Thus the decoherence along the $S_{1}$ direction followed by the
above transformation is
\begin{equation}\label{decoh4}
\pmatrix{\cosh\rho + (\sinh\rho)\cos(2\theta) &
(\sinh\rho)\sin(2\theta) \cr (\sinh\rho)\sin(2\theta) &
\cosh\rho - (\sinh\rho)\cos(2\theta) }
\pmatrix{e^{\lambda} & 0 \cr 0 & e^{-\lambda}} .
\end{equation}
The computation of this matrix algebra leads to another decoherence
matrix preceded by a rotation matrix
\begin{equation}\label{decoh5}
\pmatrix{\cosh\xi + (\sinh\xi)\cos(2\alpha) &
(\sinh\xi)\sin(2\alpha) \cr (\sinh\xi)\sin(2\alpha) &
\cosh\xi - (\sinh\xi)\cos(2\alpha) }
\pmatrix{\cos\phi & -\sin\phi \cr \sin\phi & \cos\phi} ,
\end{equation}
where
\begin{eqnarray}
\cosh\xi &=& (\cosh\lambda)\cosh\rho +
(\sinh\lambda)(\sinh\rho)\cos\theta ,  \nonumber \cr \\[1mm]
\tan\alpha &=& \frac{(\sin{\theta})[\sinh\rho + (\tanh\lambda)
(\cosh\rho - 1)\cos\theta]}{(\sinh\rho)\cos\theta + (\tanh\lambda)
[1 + (\cosh\rho - 1)(\cos\theta)^{2}} ,  \nonumber \cr \\[2mm]
\tan\phi &=& \frac{(\tanh\rho)(\tan\lambda)\sin2\theta}{1 +
(\tanh\lambda)(\tan\rho)\cos2\theta} .
\end{eqnarray}
The calculation leading to the above expression is well known from
the squeezed state and the Lorentz transformation.
The overall decoherence factors for Eq.(\ref{decoh2}) and
Eq.(\ref{decoh3}) are $e^{-\lambda}$ and $e^{-\rho}$ respectively.
The overall factor for Eq.(\ref{decoh5}) is $e^{-\xi}$, and the
net decoherence effect on $S_{B}$ is
\begin{equation}
e^{-\xi}\pmatrix{\cosh\xi + (\sinh\xi)\cos(2\alpha) &
(\sinh\xi)\sin(2\alpha) \cr (\sinh\xi)\sin(2\alpha) &
\cosh\xi - (\sinh\xi)\cos(2\alpha) }
\pmatrix{\cos\phi & -\sin\phi \cr \sin\phi & \cos\phi} .
\end{equation}
\narrowtext
The non-trivial aspect of this calculation is the rotation matrix in
the above expression.  The decoherence followed by another decoherence
does not always result in a decoherence.  It is a decoherence preceded
by a rotation.  It is a simple matter to detect this rotation once
the decoherence filters are built in laboratories.

This effect of the Lorentz group has been discussed in connection with
polarization optics~\cite{chiao88,kitano89}.  In special relativity,
this extra rotation is called the Thomas effect and manifests itself
in the energy spectrum of the hydrogen atom~\cite{corben60}.

\section{Further Physical Implications}\label{further}
We have thus far reformulated the existing mathematical devices for
polarization optics in terms of the two-by-two and four-by-four
representations of the Lorentz group.  In so doing, we have achieved
a unified group theoretical formulation of polarization optics.  The
next question then is whether this new formulation will lead to new
applications or new experiments.  This question is not unlike the
question arising from Maxwell's formulation of electromagnetism.
After putting together various aspects of electricity and magnetism
into a single mathematical formalism, we are led to the question of
whether the formalism leads to a new physics.  In the case of
Maxwell's equations, the new physics led to wireless communication
and electronic industry.

The result of this paper is not as far-reaching as in Maxwell's case,
but we are working within the same philosophical framework as the
case of Maxwell's equations.  Yes, we have unified various aspects
of polarization optics into a single group theoretical formalism
We are now interested in new conclusions which can be derived and
which can be observed in laboratories.  For this purpose, we note
that the Lorentz group has interesting subgroups.  The Lorentz
group has six generators forming a closed set of commutation
relations.   We have already used this concept when we discussed
a system consisting only of phase shifters, which is governed by
$O(3)$ or the three-dimensional rotation group with three generators.
The group $O(3)$ is a subgroup of the Lorentz group.  It was noted
that the system of attenuators is governed by the $O(2,1)$ subgroups
of the Lorentz group again with three generators.

We have achieved the full six-parameter Lorentz group by combining
both the phase-shifters and attenuators.  We are then led to the
question of whether this full Lorentz group has subgroups other than
the input groups $O(3)$ and $O(2,1)$.  As we discuss in Appendix A,
there is an interesting subgroup which is like the two-dimensional
Euclidean group.  This is the product of our group theoretical
formulation.  In this section, we outline first the result of our
effort on the Jones-matrix formalism~\cite{hkn97josa}.  We shall then
extend this result to the case of Stokes parameters.

The Lorentz group has three boost and three rotation generators.  As
we shall note in Appendix A, we can construct a set of generators
consisting of $J_{3}$, $N_{1}$ and $N_{2}$, with
\begin{equation}\label{n12}
N_{1} = J_{1} + K_{2} , \qquad   N_{2} = J_{2} - K_{1} .
\end{equation}
These generators satisfy the commutation relations:
\begin{equation}\label{e2com}
[J_{3}, N_{1}]  = i N_{2} , \quad [J_{3}, N_{2}] = -iN_{1} , \quad
[N_{1}, N_{2}]  = 0 .
\end{equation}
In the case of two-by-two Jones-matrix formalism, they take the form
\begin{equation}
N_{1} = \pmatrix{0 & 1 \cr 0 & 0},  \qquad
N_{2} = \pmatrix{0 & -i \cr 0 & 0} .
\end{equation}
They indeed form a closed set of commutation relations.  As shown in
Appendix B, these commutation relations are like those for the
two-dimensional Euclidean group consisting of two translations and one
rotation around the origin.  This group has been studied extensively in
connection with the space-time symmetries of massless particles,
where $J_{1}$ and the two $N$ generators correspond to the helicity and
gauge degrees of freedom respectively~\cite{hks82}.

However, this group is relatively new in optics~\cite{hkn97josa,ky92},
and we are tempted to construct an optical filter possessing this symmetry.
The physics of $J_{3}$ is well known through the phase shifter given in
Eq.(\ref{shif1}).  If the angle $\delta$ is $\pi/2$, the phase shifter
becomes a quarter-wave shifter, which we write as
\begin{equation}
Q = P(0, \pi/2) = \pmatrix{e^{-i\pi/4} & 0 \cr 0 & e^{i\pi/4}} .
\end{equation}
Then $J_{1}$ and $K_{2}$ are the quarter-wave conjugates of $J_{2}$ and
$K_{1}$ respectively:
\begin{equation}
J_{1} = Q J_{2} Q^{-1} , \qquad K_{2} = - QK_{1}Q^{-1} .
\end{equation}
Consequently,
\begin{equation}
N_{1} = Q N_{2} Q^{-1} .
\end{equation}
The $N$ generators lead to the following transformation matrices.
\begin{eqnarray}\label{tmatrix}
T_{1}(u) &=& \exp{(-iu N_{1})}
= \pmatrix{1 & iu \cr 0 & 1}, \nonumber \\[2mm]
T_{2}(v) &=& \exp{(-iv N_{2})} = \pmatrix{1 & -v \cr 0 & 1} .
\end{eqnarray}
It is clear that $T_{1}$ is the quarter-wave conjugate of $T_{2}$.
We can now concentrate on the transformation matrix $T_{2}$.

If $T_{2}$ is applied to the incoming wave of Eq.(\ref{expo1}),
\begin{equation}\label{tau}
\pmatrix{1 & -v \cr 0 & 1} \pmatrix{E_{x} \cr E_{y}} =
\pmatrix{E_{x} - v E_{y} \cr E_{y}} .
\end{equation}
This new filter superposes the $y$ component of the electric field to
the $x$ component with an appropriate constant, but it leaves the $y$
component invariant.

Let us examine how this is achieved.  The generator $N_{2}$ consists
of $J_{2}$ which generates rotations around the $z$ axis, and $K_{1}$
which generates a squeeze along the $45^{o}$ axis.  Physically, $J_{2}$
generates optical activities.  Thus, the new filter consists of a
suitable combination of these two operations.  In both cases, we have
to take into account the overall attenuation factor.   This can be
measured by the attenuation of the $y$ component which is not affected
by the symmetry operation of Eq.(\ref{tau}).

Is it possible to produce optical filters of this kind?  Starting from
an optically active material, we can introduce an asymmetry in absorption
to it by either mechanical or electrical means.  Another approach would
be to pile up alternately the $J_{3}$-type and $K_{2}$-type layers.  In
either case, it is interesting to note that the combination of these two
effects produces a special effect predicted from the Lorentz group.

The group $E(2)$, although new in optics, has many interesting
properties having to do with our daily life.  One important property
is the conversion of multiplication into addition as the following
matrix algebra indicates:
\begin{equation}
\pmatrix{1 & v_{1} \cr 0 & 1}  \pmatrix{1 & v_{2} \cr 0 & 1}
= \pmatrix{1 & v_{1} + v_{2} \cr 0 & 1} .
\end{equation}
Since this group deals with rotations and translations on a plane,
it has a great potential in navigational sciences.  However, we are
here interested in what role this group plays in the Stokes parameters.

In the Jones-matrix formalism, we used two-by-two matrices for
transformations.  For the Stokes parameters, we have to use four-by-four
matrices applicable to Stokes four-vectors.  The four-by-four
generators of the Lorentz transformations are given in
Sec.~\ref{trstokes}.  They are discussed in more detail in Appendix A.
There, the generators $N_{1}$ and $N_{2}$ are derived from the boost
and rotation generators.  From these generators, we can construct
transformation matrices:
\begin{eqnarray}
T_{1}(u) &=& \exp{(-iuN_{1})}  \nonumber \\
&=& \pmatrix{1 + u^{2}/2 & 0 & u & -u^{2}/2 \cr 0 & 1 & 0 & 0 \cr
u & 0 & 1 & -u \cr u^{2}/2 & 0 & u & 1 - u^{2}/2} , \nonumber \\[2mm]
T_{2}(v) &=& \exp{(-ivN_{2})} \nonumber \\
&=& \pmatrix{1 + v^{2}/2 & -v & 0 & -v^{2}/2 \cr -v & 1 & 0 & v \cr
0 & 0 & 1 & 0 \cr v^{2}/2 & -v & 0 & 1 - v^{2}/2} .
\end{eqnarray}
These expressions are the four-by-four representation of the
transformation matrices given in Eq.(\ref{tmatrix}) for the Jones
spinors.

Unlike the Jones-matrix formalism, the Stokes parameters can describe
partially coherent light waves.  This is the reason why the above
four-by-four expression is complicated.  These transformation matrices
preserve coherence.  If decoherence is introduced, we can apply the
decoherence matrices discussed in Sec.~\ref{deco}.  It is interesting
to note that the Lorentz-group formulation of polarization optics opens
up this kind of new possibilities in physics.

\section*{Concluding Remarks}
In this paper, we have shown that both the Jones-matrix formalism and
the Stokes parameters can be formulated as two different representations
of the same Lorentz group.  The physics of the Stokes parameters
can deal with the coherence between the two polarization directions.
It is shown also that the decoherence effect can also be formulated
within the framework of the Lorentz group or in terms of the
mathematics of squeezed states of light.

There have been in the past many laudable attempts to construct a
mathematical representation for polarization optics based on the
Lorentz group~\cite{barakat63,parent60,pellat91}.  However, the Lorentz
group, particularly its relevance to optics, was not fully appreciated
until it started playing the role of the underlying symmetry group
for squeezed states of light~\cite{knp91,dir63,yuen76,cav85}.
This naturally led to a new interest in possible applications of the
Lorentz group in other branches of optics including polarization
optics~\cite{chiao88,kitano89,pellat91}.

From the group theoretical point of view, what is new in this paper
is that we used in Sec.~\ref{spinf} an additional symmetry of the
Lorentz group to understand fully the connection between the Jones
matrix and the Stokes parameters.  This additional symmetry was
the one which connects electrons with positrons through charge
conjugation.  This opens a new research line which will connect
symmetries of relativistic particles with polarization optics.
We can attempt to understand the symmetries of particle physics
not from commutation relations of group generators but from what
we observe in optics laboratories.

It is true that we used group theory as the main carrier of our
analysis.  On the other hand, it is true also that we did not
start our paper with commutation relations, but with what we
observe in the real world.  We concluded this paper with what
we can observe or we may possibly observe in the real world.

\appendix

\section{Subgroups of the Lorentz Group}
Let us consider the space-time coordinates $(t, x, y, z)$, analogous
to the Stokes parameters $(S_{0}, S_{1}, S_{2}, S_{3})$.  Then the
rotation around the $z$ axis is performed by the four-by-four matrix
\begin{equation}
\pmatrix{1 & 0 & 0 & 0 \cr 0 & \cos\theta & -\sin\theta & 0  \cr
0 & \sin\theta & \cos\theta & 0 \cr 0 & 0 & 0 & 1} .
\end{equation}
This transformation is generated by $J_{3}$ of Eq.(\ref{jj31}).
The generators of rotations around the $x$ and $y$ axes are also given
in Eq.(\ref{jj31}) and Eq.(\ref{jj2}).  These three generators
satisfy the closed set of commutations relations
\begin{equation}\label{rota}
\left[J_{i}, J_{j} \right] = i \epsilon_{ijk} J_{k} .
\end{equation}
This set of commutation relations is for the three-dimensional rotation
group.

The Lorentz boost along the $z$ axis takes the form
\begin{equation}
\pmatrix{\cosh\eta & 0 & 0 & \sinh\eta \cr 0 & 1 & 0 & 0 \cr
0 & 0 & 1 & 0 \cr \sinh\eta & 0 & 0 & \cosh\eta} ,
\end{equation}
which is generated by $K_{3}$ of Eq.(\ref{kk23}).
Boosts along the $x$ and $y$ axes are generated by $K_{1}$ and $K_{2}$
given in Eq.(\ref{kk1})  and Eq.(\ref{kk23}) respectively.
These boost generators satisfy the commutation relations
\begin{equation}\label{boosta}
\left[J_{i}, K_{j} \right] = i \epsilon_{ijk} K_{k} , \qquad
\left[K_{i}, K_{j} \right] = -i \epsilon_{ijk} J_{k} .
\end{equation}

Indeed, the three rotation generators and the three boost generators
satisfy the closed set of commutation relations given in Eq.(\ref{rota})
and Eq.(\ref{boosta}).  The four-by-four transformation matrices
generated by these generators are directly applicable to the
space-time four-vector $(t, x, y, z)$ and also to the Stokes four-vector
$(S_{0}, S_{1}, S_{2}, S_{3})$.

We can now construct a subset of the generators consisting of
$J_{3}$, $N_{1}$ and $N_{2}$ defined in Eq.(\ref{n12}).  $J_{3}$ is
given in Eq.(\ref{jj31}).  Thus the generators $N_{1}$ and
$N_{2}$ take the form
\begin{equation}
N_{1} = \pmatrix{0 & 0 & i & 0 \cr 0 & 0 & 0 & 0 \cr
i & 0 & 0 & -i \cr 0 & 0 & i & 0},  \quad
N_{2} = \pmatrix{0 & -i & 0 & 0 \cr -i & 0 & 0 & i \cr
0 & 0 & 0 & 0 \cr 0 & -i & 0 & 0} .
\end{equation}
These four-by-four matrices satisfy the closed set of commutation
relations given in Eq.(\ref{e2com}).

The subgroup of the Lorentz group generated by the above matrices
governs the internal space-time symmetry of massless particles, and
has been extensively discussed in the
literature~\cite{hks86,knp86,wein64}.  These expressions are new in
optics.

\section{Two-dimensional Euclidean Transformations}
In Sec.~\ref{further}, we discussed a set of commutation relations
satisfied by the generators of the two-dimensional Euclidean group.
The purpose of this Appendix is to construct the generators for
the group of transformations on a flat plane.  There are translations
and rotations.

Let us consider here a two-dimensional plane and use the $xy$ coordinate
system.  Then $L_{z}$ defined as
\begin{equation}
L_{z} = - i\left\{x {\partial \over \partial y} -
y {\partial \over \partial x} \right\}
\end{equation}
will generate rotations around the origin.  The translation generators
are
\begin{equation}
P_{x} = -i {\partial \over \partial x} , \qquad
P_{y} = -i {\partial \over \partial y} .
\end{equation}
These generators satisfy the commutation relations:
\begin{equation}
[L_{z}, P_{x}] = i P_{y},  \qquad [L_{z}, P_{y}] = -iP_{x} \qquad
[P_{x}, P_{y}] = 0 .
\end{equation}
These commutation relations are like those given in Eq.(\ref{e2com}).
They become identical if $L_{z}$, $P_{x}$ and $P_{y}$ are replaced by
$J_{1}$, $N_{2}$ and $N_{3}$ respectively.

This group is not discussed often in physics, but is intimately related
to our daily life.  When we drive on the streets, we make translations
and rotations, and thus make transformations of this $E(2)$ group.  In
addition, this group reproduces the internal internal space-time
symmetry of massless particles~\cite{wig39}.  This aspect of the $E(2)$
group has been extensively discussed in the
literature~\cite{hks86,knp86,wein64,hks82}.

\end{document}